**Highlights**

- The effective Schrödinger equation for pendulum with fast suspension point vibration
- The Kapitza pendulum quantum states
- Models for stabilization potential
- The state energy correction due to the tunnel penetration of the barrier between wells

# Quantum states of the Kapitza pendulum

P.A. Golovinski, V.A. Dubinkin

Voronezh State Technical University, Voronezh, Russia

golovinski@bk.ru

The quantum states of the Kapitza pendulum are described within the effective potential obtained by the method of averaging over the fast oscillations. An analytical estimate of the energy spectrum of stabilized states is given using approximate model potential. For the lowest states of an inverted pendulum, the spectrum is repeduced by the energies of a harmonic oscillator with perturbation theory corrections. Tunneling effect contribution to the energies of resonance states in the double-well effective potential is estimated. The results of numerical simulations of vibrational and rotational spectra of the Kapitza pendulum by the semiclassical method and by the Numerov algorithm are compared.

## 1. Introduction

Starting by the Kapitsa seminal systematic investigation [1] of unusual features of equilibrium of a rigid pendulum with a vibrating suspension point [2], the phenomenon of its stabilization in the upper position has been studied in details [3]. This effect, explained with the analytical Kapitza method [4], based on the asymptotic separation of fast and slow variables [5], has found wide application in various branches of classical and quantum physics. The dynamics of nonlinear classical mechanical systems in high-frequency fields can be described by an effective time-independent Hamiltonian. This Hamiltonian is obtained by successive expansion of the solution in inverse powers of frequency and allows one to study motion in the presence of rapidly oscillating fields within the framework of the theory developed for autonomous systems [6]. A nonlinear system can be stabilized also by noise, so that the upper equilibrium point of the pendulum becomes stable even when the noise is white despite the simple Kapitza pendulum conditions are not fulfilled [7]. A stable vertical position of an inverted pendulum also observed when suspension point is subjected to the combined action of high-frequency harmonics and stochastic forces [8]. The stabilization effect due to vibration was extended to complex elastic systems [9-12], and the application of this idea to the motion of charged particles in electromagnetic waves led to the concept of the Miller force [13, 14] as well gave the optical tweezer principle [15-17].

Phenomena similar to the Kapitza effect manifests itself in non-mechanical systems, for example, when considering the nonlinear Schrödinger equation with a periodically varying dispersion coefficient for pulse stabilization in fiber-optic systems [18], and in structures with a transverse distribution of the refractive index, been periodically modulated along the longitudinal coordinate leading to light confinement [19, 20]. The stabilization theory based on the averaging method can be generalized to the imagine oscillating potential. At a high frequency, a bound state is formed in it, which can ensure the stability of an optical resonator with variable reflectivity [21], and due to amplitude and phase modulation, stabilization of the unstable mode of resonator solitons is achieved [22].

Quantum dynamics in a high-frequency field, like the classical one, is described by an effective time-independent Hamiltonian [23]. This effective Hamiltonian determines the states, as well as the magnitude of the lowest resonance in the atomic trap model [24], the experimental implementation of which is realized by combining a static configuration with an external alternating electric field [25]. Such an oscillating field captures particles, since at least one bound state always exists in the one-dimensional well of the effective potential [26, 27]. Experimental installations with ultracold atoms built on this principle are convenient for studying quantum systems far from equilibrium. Based on these instruments, it is possible to produce the Floquet substance with the Kapitza stabilization effect for trapped atoms at the maxima of the optical lattice field [28–30].

Dynamic Kapitza stabilization in many-particle systems prevents heating beyond a certain threshold of periodic action [31] and causes a number of other phenomena. In particular, the periodic modulation of the transverse magnetic field makes it possible to ensure stable trapping of ferromagnetic spin systems around unstable paramagnetic configurations [32]. A generalization of the Kapitza pendulum to a system of many bodies is the sine-Gordon model with a periodic action, dynamically stable when exposed to a finite frequency and amplitude [33]. In the controlled two-mode Bose-Hubbard model in the Josephson mode for a weak nonresonant interaction and small chaotic component the collective behavior also reproduces stabilization of the Kapitza pendulum [34].

The wide manifestation of the Kapitza stabilization effect in various quantum systems dictates a quantum consideration of this phenomenon. The Kapitza quantum pendulum is stabilized in the form of quantum states near a local minimum of the effective potential energy. All the quantum states of the Kapitza pendulum need to be properly described, and we consider quantum states in such a potential to determine their energies.

## 2. The quantum equation of motion of the Kapitza pendulum

We will discuss the Kapitza pendulum, i.e. a plane pendulum with a length $l$ in a uniform field, the suspension point of which performs high-frequency vertical oscillations according to the dependence $a \cos \omega t$ [5] with amplitude $a << l$. The coordinates of a point with mass $m$ are

$$x = l \sin \theta, \; y = a \cos \omega t + l \cos \theta, \qquad (2.1)$$

and the Lagrange function

$$L = \frac{J}{2} \dot{\theta}^2 + m l a \omega^2 \cos \omega t \cos \theta + m g l \cos \theta, \qquad (2.2)$$

where $J = m l^2$. The force $f = -m l a \omega^2 \cos \omega t \sin \theta$ is assumed to be rapidly variable with the frequency

$$\omega >> \sqrt{g/l} . \qquad (2.3)$$

Then the classical equation of motion has the form

$$\ddot{\theta} + l^{-1}(g + a \omega^2 \cos \omega t) \sin \theta = 0, \qquad (2.4)$$

and after performing averaging over fast oscillations, the effective potential energy is obtained as

$$U_{eff}(\theta) = U \left( 2\alpha \sin^2 \theta - \cos \theta \right), \qquad (2.5)$$

where $U = mgl$, $\alpha = \dfrac{a^2 \omega^2}{8gl}$. The point $\sin \theta = 0$ corresponds to the local minimum of the function $U_{eff}(\theta)$. The condition $\alpha > 1/4$ provides stability of the upper position $\theta = \pi$ with $U_{eff}(\pi) = U$. The global minimum $-U$ of potential energy $U_{eff}(\theta)$ is placed at $\theta = 0$.

The Schrödinger equation for the wave function of the Kapitza quantum pendulum has the form

$$i\hbar \frac{\partial \Psi}{\partial t} = -\frac{\hbar^2}{2J} \frac{\partial^2 \Psi}{\partial \theta^2} - ml\left(a\omega^2 \cos\omega t \cos\theta + g\cos\theta\right)\Psi. \qquad (2.6)$$

We use the Cook transform [35] for the wave function

$$\Psi(\theta,t) = \psi(\theta,t)\exp\left(-i\frac{mla\omega^2 \cos\theta}{\hbar\omega}\sin\omega t\right) \qquad (2.7)$$

for further excluding the fast oscillations. After substituting expression (2.7) into equation (2.6) and averaging over the period of fast oscillations of the external force, we obtain the Schrödinger equation with the effective potential in the form

$$i\hbar \frac{\partial \psi}{\partial t} = -\frac{\hbar^2}{2J} \frac{\partial^2 \psi}{\partial \theta^2} + U_{eff}(\theta)\psi. \qquad (2.8)$$

Stationary states in a time-independent one-dimensional effective potential exist at its any parameters, and the equation for these states has the form

$$\frac{d^2\phi}{d\theta^2} + \frac{2J}{\hbar^2}\left(E - U_{eff}(\theta)\right)\phi = 0. \qquad (2.9)$$

This is so-called Whittaker-Hill equation with studied general properties [36, 37]. In the limit $E \gg U$, equation (2.9) describes a plane rigid quantum rotator with normalized periodic solution

$$\phi = \frac{1}{\sqrt{2\pi}}\exp(\pm in\theta), n = 0,1,2,... \qquad (2.10)$$

and discrete energy spectrum counted from the bottom of the potential well

$$E_n = \frac{\hbar^2 n^2}{2J} - U. \qquad (2.11)$$

The quantum states of a plane pendulum in a uniform field without suspension vibration ($f = 0$) are well-known [38]. Like the solutions for free rotation, such states have completely discrete energy spectrum that has a topological origin. In contrast to this, the energies of one-dimensional free particle motion have a continuous spectrum, as do the energies in the infinite motion mode in a one-dimensional infinitely extended potential of the equation (2.5) kind. We father consider the motion modes for the quantum Kapitza pendulum with a vibrating suspension, using measurement units, in which $U = 1, \hbar = 1$.

### 3. Vibrational states

The effective potential for the Kapitza pendulum has a form of one stabilizing shallow well and another deep potential well, separated by a potential barrier. The tops of the symmetric maxima $U_{max}$ of potential energy are located at the points

$$\theta_{max} = \pi \mp \arccos(1/4\alpha), \qquad (3.1)$$

and

$$U_{max} = 2\alpha + \frac{1}{8\alpha}. \qquad (3.2)$$

To make analytical estimate of the energies for the shallow well states, we replace the effective potential (2.5) in the vicinity of point $\theta = 0$ by the model potential

$$U_{well}(\theta) = -\frac{U_{max} - 1}{\cosh^2 k_{min}(\theta - \theta_{min})} + U_{max}. \qquad (3.3)$$

The energy spectrum of the bound states in potential (3.3) is described by the formula [26]

$$E_n - U_{max} = -W\left[\sqrt{1 + \frac{U_{max} - 1}{W}} - (1 + 2n)\right]^2, \qquad (3.4)$$

where $W = (8J)^{-1}k_{min}^2$. The condition for the existence of only one bound state has the form

$$U_{max} - 1 < k_{min}^2 / J. \qquad (3.5)$$

Near point $\theta_{min}$ of the effective potential bottom for a well, the potential is close to the quadratic function of the deviation angle, and $U_{eff}(\theta)$ is approximately replaced to

$$U_{osc}(\theta) = \pm 1 + \frac{C(\theta - \theta_{min})^2}{2}, C = 4\alpha \mp 1, \qquad (3.6)$$

where the upper sign corresponds to a shallow well, and the lower sign corresponds to a deep well. Equation (2.9) takes the form

$$\frac{d^2\phi}{d\theta^2} + 2J\left(E \mp 1 - \frac{C(\theta - \theta_{min})^2}{2}\right)\phi = 0. \qquad (3.7)$$

We introduce the notation $\xi = \mu(\theta - \theta_{min})$, $\mu = (JC)^{1/4}$ and $\lambda = 2(E \mp 1)/\omega_c$, where $\omega_c = ((4\alpha \mp 1)/J)^{1/2}$ is oscillation frequency of a classical oscillator. In these notations, equation (3.7) is write down as

$$\frac{d^2\phi}{d\xi^2} + \left(\lambda - \xi^2\right)\phi = 0. \qquad (3.8)$$

Its solutions are expressed in terms of Hermite polynomials of order $n$ in the form

$$\phi_n(\theta) = \left(\frac{\mu^2}{\pi}\right)^{1/4} \frac{1}{\sqrt{2^n n!}} H_n(\xi) \exp(-\xi^2/2), \qquad (3.9)$$

and energies are $E_n = 1 + (n+1/2)\omega_c$. Such energies systematically overestimate the true values, since the oscillator potential is narrower than the effective potential.

To obtain more accurate energy values, we use the perturbation theory in the basis of functions (3.9). The states of the quantum oscillator are not degenerate; therefore, the nonzero correction to the energy in the first order with respect to the perturbation $H_I(\theta) = U_{eff}(\theta) - U_{osc}(\theta)$ is

$$E_n^{(1)} = \langle n|H_I|n\rangle = \langle n|U_{eff}|n\rangle - 1 - \frac{1}{2}(n+1/2)\omega_c. \qquad (3.10)$$

After evaluating the matrix element $\langle n|U_{eff}|n\rangle$ we get

$$E_n^{(1)} = \alpha - 1 - \frac{1}{2}(n+1/2)\omega_c + \exp(-1/4\mu^2)L_n(1/2\mu^2) - \alpha(\exp(-1/\mu^2)L_n(2/\mu^2), (3.11)$$

where $L_n(x)$ is Laguerre polynomials. Formula (3.11) allows calculating the energies of the lower levels, both in the shallow well and in the deep well.

## 4. Semiclassical description

The number of states grows with $J$ increasing and the behavior of the system tends to the classical one. We consider now the bound states in the double-well potential in the semiclassical approximation [39]. Stable classical motion of an inverted pendulum with angular momentum

$$p(E,\theta) = \sqrt{2J(E - U_{eff}(\theta))} \qquad (4.1)$$

in the potential (2.5) is permitted between the classical turning points in a shallow well where $p(E,\theta_{1,2}) = 0$. Between the boundaries of the potential barrier $\theta_2$ and $\theta_3$, the motion is classically forbidden, and penetration through the barrier is moderated by quantum tunneling effect. The solution to the equation $E - U_{eff}(\theta) = 0$ determines the position of the barrier turning points

$$\cos\theta_{2,3} = -\frac{1}{4\alpha} \pm \sqrt{1 - \frac{E}{2\alpha} + \frac{1}{16\alpha^2}}. \qquad (4.2)$$

When $E = U_{\max}$, the turning points merge into one, the potential barrier disappears, and the stabilization is absent.

Semiclassical quantization rule to determine the energy $E_n$ in a potential well, we write down in the form

$$S_{12} = (n + 1/2)\pi, \qquad (4.3)$$

where phase integral $S_{12} = S(E, \theta_1, \theta_2) = \int_{\theta_1}^{\theta_2} p(E, \theta)\, d\theta$ and $n = 0,1,2....$ For a finite motion, the maximum value of the phase integral $S_{12}$ is reached when the classic turning points are located at the symmetrical tops $U_{max}$. Quantum effects are most noticeable when $\sqrt{2J(U_{max} - U_{min})} \sim 1$. With an increase in this parameter, the energy spectrum gradually turns into a continuous one, and the motion is less and less different from the classic.

In two adjacent wells of the Kapitsa pendulum, states with close energies can exist. Then, the resonant tunneling passage of the pendulum from one well to another ensures the formation of two bilocalized states. To find their energy, we use the transformation matrix of solutions for the period [40]. The periodic Floquet solution to the Hill equation satisfies the condition

$$\lambda\phi(\theta + 2\pi) = \phi(\theta). \qquad (4.4)$$

The transformation matrix of the wave function for a considered potential is obtained in the semiclassical approximation [41]. To find the energies of stationary states, we choose in the interval $\theta_1 < \theta < \theta_2$ two linearly independent solutions [42]

$$u = p^{-1/2}\exp\left(iS(E, \theta_1, \theta)\right), v = u^*. \qquad (4.5)$$

The transformation of these functions over the period will take the form

$$u(\theta + 2\pi) = Du(\theta) + Rv(\theta), \qquad (4.6)$$

$$v(\theta + 2\pi) = D^*v(\theta) + R^*u(\theta). $$

Condition (4.4) leads to the equation

$$\begin{vmatrix} D - \lambda & R \\ R^* & D^* - \lambda \end{vmatrix} = 0. \qquad (4.7)$$

Taking this into account, the spectral equation takes the form

$$\lambda^2 - 2\operatorname{Re}D\lambda + 1 = 0, \qquad (4.8)$$

because $DD^* - RR^* = 1$ due to the Wronskian's properties, and

$$\operatorname{Re}D = \frac{1}{2}\beta\cos S_{12}\cos S_{34} - \sin S_{12}\sin S_{34}, \qquad (4.9)$$

where $\beta = 4\sigma^2 + \dfrac{1}{4\sigma^2}$ , $\sigma(E) = \exp\left(\dfrac{1}{\hbar}\int\limits_{\theta_2}^{\theta_1}|p(E,\theta)|d\theta\right)$ . Since for the stationary states on a

circle, the multiplier is $\lambda = 1$ ., the condition determining the energies takes the form

$$\operatorname{Re} D = 1 . \qquad (4.10)$$

If the energies of the states in the wells differ more than $\beta^{-1}$, these states are practically independent. For closely spaced levels, their mutual influence can be taken into account as a perturbation, considering the action as a small deviation $\delta_{ij}$ the phase from the quantization condition

$$S_{ij} = (n + 1/2)\pi + \delta_{ij} . \qquad (4.11)$$

Here $n = n_1$ for a shallow well and $n = n_2$ for a deep well. Using representation (4.9), taking into account Eq. (4.10), we get

$$\frac{1}{2}\beta\delta_{12}\delta_{34} = 1 + (-1)^{n_1 + n_2} . \qquad (4.12)$$

If the quantum numbers $n_1$ and $n_2$ have different parities, then in the two sub-barrier regions the signs of the interaction of the states are opposite, and there is no energy splitting. For the states with the same parity

$$\frac{dS_{12}}{dE}\frac{dS_{34}}{dE}(\Delta E)^2 = \frac{4}{\beta} . \qquad (4.13)$$

Given that $dS_{ij}/dE = \pi/\omega_{ij}$, the splitting of the energies of states will be expressed in the form

$$\Delta E = \pm\frac{2}{\pi}\sqrt{\frac{\omega_{12}\omega_{34}}{\beta}} , \qquad (4.14)$$

where $\omega_{12}$ is classical oscillation frequency in a shallow well, $\omega_{34}$ is classical oscillation frequency in a deep well. For $\sigma \gg 1$ we have

$$\Delta E = \pm\frac{\sqrt{\omega_{12}\omega_{34}}}{\pi}\sigma^{-1} \qquad (4.15)$$

that is twice the value of the tunneling splitting of the energy levels found for an asymmetric double well, both in the semiclassical approximation [43] and by the method of two-level approximation [44], due to the twice subbarrier overlap of wave functions.

For the over-barrier rotational motion of the Kapitsa pendulum with energy $E > U_{max}$ , the condition for the applicability of the procedure for the wave function averaging over fast oscillations of the external field takes the form

$$\omega >> \omega_r, \qquad (4.16)$$

where $\omega_r$ is classical frequency of a pendulum rotation. The semiclassical Bohr-Sommerfeld quantization condition for rotational motion in the effective potential is written as

$$S(E, 0, 2\pi) = 2\pi n. \qquad (4.17)$$

Equation (4.17) allows one to find the energies of states, allowing for the influence of the effective potential and the motion periodicity.

## 5. Results of numerical simulations and discussions

As an example illustrating a general theoretical description, we take a Kapitza pendulum with parameters $\alpha = 1/2$ and $J = 270$. This choice of parameters ensures several quantum levels in the shallow well. The effective potential energy and the corresponding model potential (3.3) with $\kappa_{min} = 1.8$ and six discrete levels are shown in Fig. 1.

**Table 1.** Energies of levels in the shallow well.

| Model potential | Semiclassical approximation | Shooting method | Harmonic oscillator | Harmonic oscillator with correction |
|---|---|---|---|---|
| 1.0358 | 1.0300 | 1.0296 | 1.0304 | 1.0296 |
| 1.1015 | 1.0875 | 1.0869 | 1.1520 | 1.0874 |
| 1.1552 | 1.1409 | 1.1403 | 1.0912 | 1.1421 |
| 1.1969 | 1.1896 | 1.1892 | 1.2128 | 1.1939 |
| 1.2266 | 1.2317 | 1.2304 | 1.2736 | 1.2429 |
| 1.2443 | | | | |

Table 1 shows the results of calculations the energy of states in the model potential, as well as in the effective potential in the semiclassical approximation, by the shooting method [45] using the numerical Numerov's algorithm [46] when integrating the Schrödinger equation, and in the harmonic oscillator approximation with a correction to the true effective potential. The wave function asymptotic for starting values of the numerical Numerov procedure is taken inside the barrier in the semiclassical approximation, which limits the accuracy of calculations by this method. The wave function of the ground state in the oscillator potential is a Gaussian function. Considering its dispersion as a parameter, the variational method found the value of the lowest state energy in a shallow well, which coincides with the calculations both by the shooting method and in the first order of the perturbation theory for a harmonic oscillator.

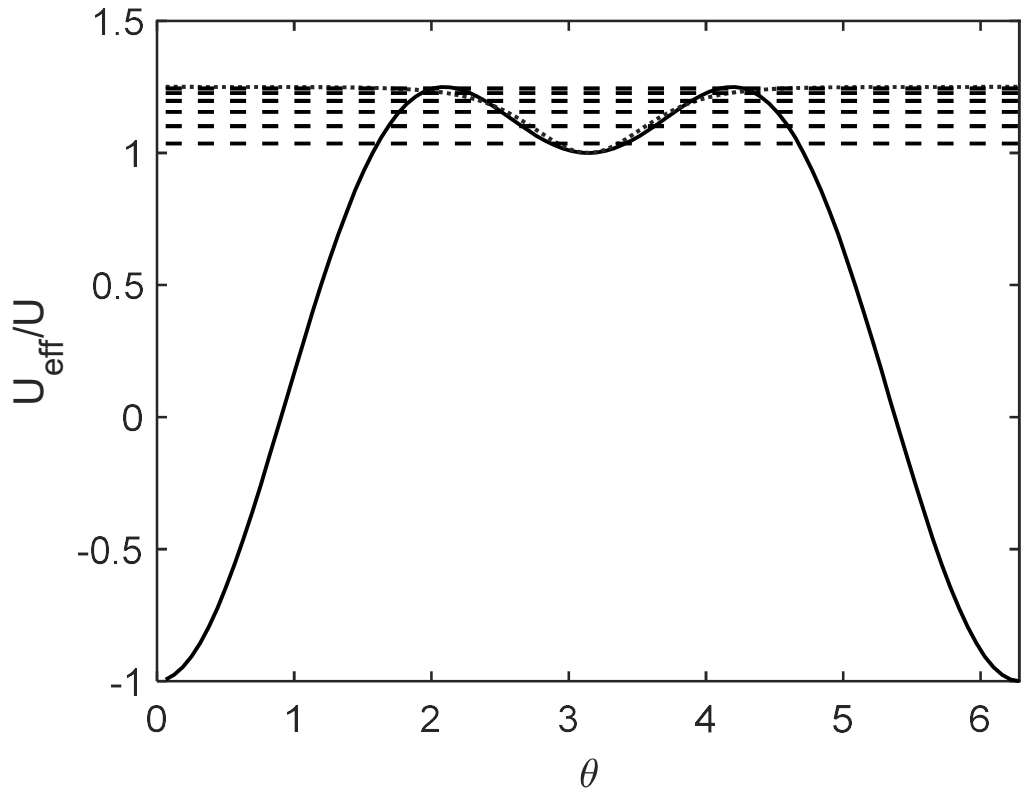

**Fig. 1.** Effective potential as a function of angle (solid line). The model potential is shown by dots, the energy levels of bound states are shown by dashed lines.

Comparison of the results for the energy spectrum obtained by different numerical simulations shows that the model potential is suitable for estimation of the energy levels position in a shallow well, that ensures the stability of an inverted pendulum, but indicates an additional state which is not reproduced by semiclassical or direct numerical calculations. This difference of exact results and the semiclassical calculations for a given model potential is explained by the known limitations of the WKB method accuracy. However, for more accurate conclusions related directly to the states in the effective potential of the Kapitza pendulum, additional research is required, taking into account the features of the behavior of the wave function near the top of the potential barrier.

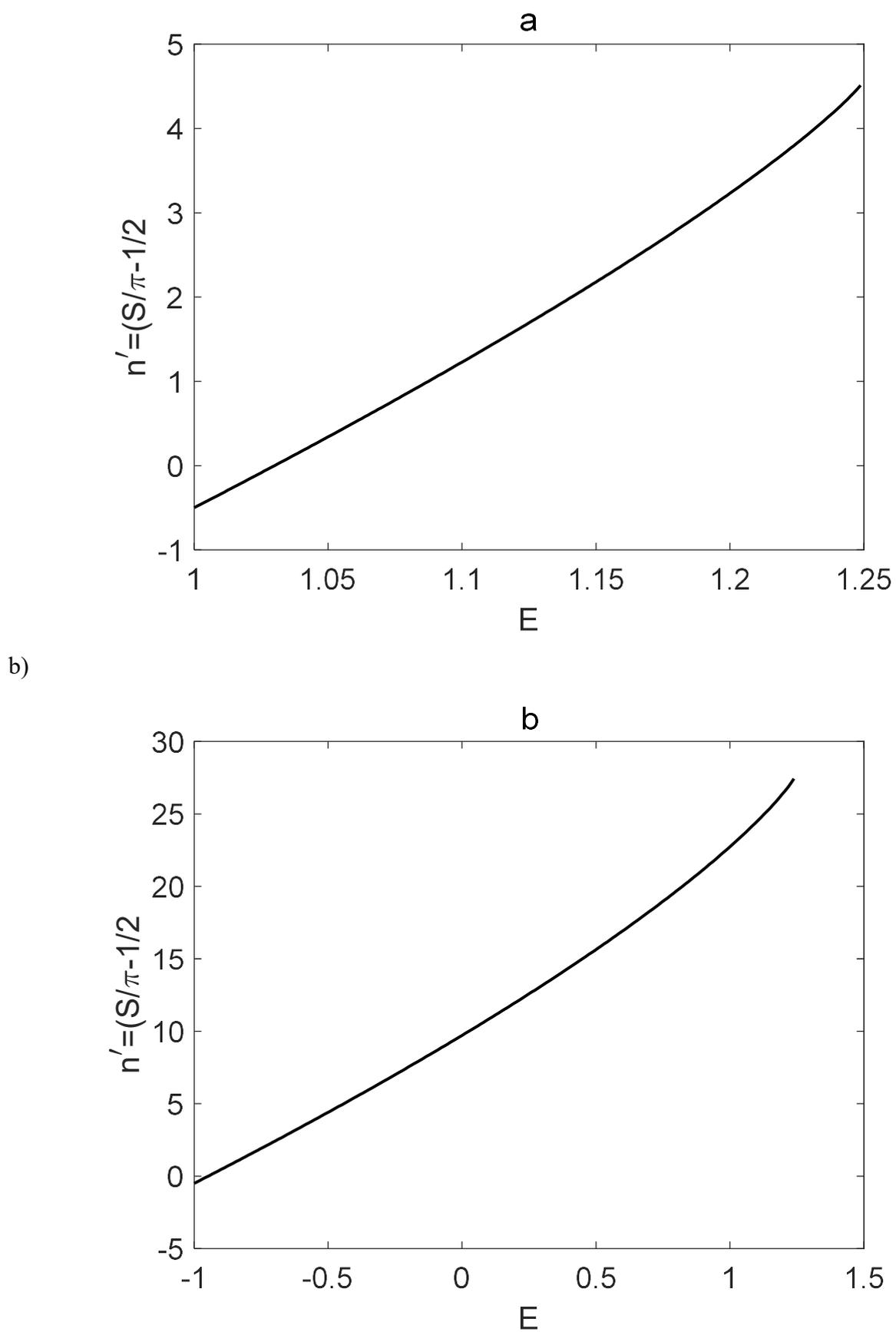

b)

**Fig. 2.** Dependence of the reduced phase $n'$ on energy for shallow well (a) and deep well (b).

Fig. 2 shows the dependence on the reduced phase $n' = (S - 1/2)/\pi$ on energy, expressed in terms of magnitude the classical action in the interval between the turning points for a shallow well and a deep well. The semiclassical energies obtained in this way at $n' = n$ correspond to 5 levels for the shallow well and demonstrate 28 states in the deep well. Five semiclassical states (from 23 to 28) in the deep well for the energy range from the bottom to the top of the shallow well potential are candidates for resonant states interaction. The uppermost states with $n_1 = 4$ and $n_2 = 27$ are most close to resonance, but, however, they have different parities, their interaction is zero, and there is no energy splitting for them due to resonant tunneling. The wave functions of these states, which have different symmetries, are calculated using the Numerov algorithm and shown in Fig. 3.

The energies of the upper states in the deep well and the energies of rotation states, calculated using the semiclassical quantization and direct numerical solution by the shooting method, are presented in Table 2. It is seen from the obtained data, as the energy increases, the accuracy of the semiclassical approximation becomes better.

**Table 2.** Energies of levels in the deep well and rotational states.

| Semiclassical approximation for a deep well | Shooting method for a deep well | Semiclassical approximation for rotation states | Shooting method for rotation states |
|---|---|---|---|
| 1.0160 | 1.0154 | 1.2619 | 1.2629 |
| 1.0743 | 1.0736 | 1.3034 | 1.2827 |
| 1.1287 | 1.1280 | 1.3537 | 1.3004 |
| 1.1786 | 1.1782 | 1.4106 | 1.3239 |
| 1.2225 | 1.2215 | 1.4735 | 1.3503 |
| | | 1.5418 | 1.3777 |
| | | 1.6152 | 1.465 |
| | | 5.182 (n=50) | 5.0699 |
| | | 7.218 (n=60) | 7.1890 |

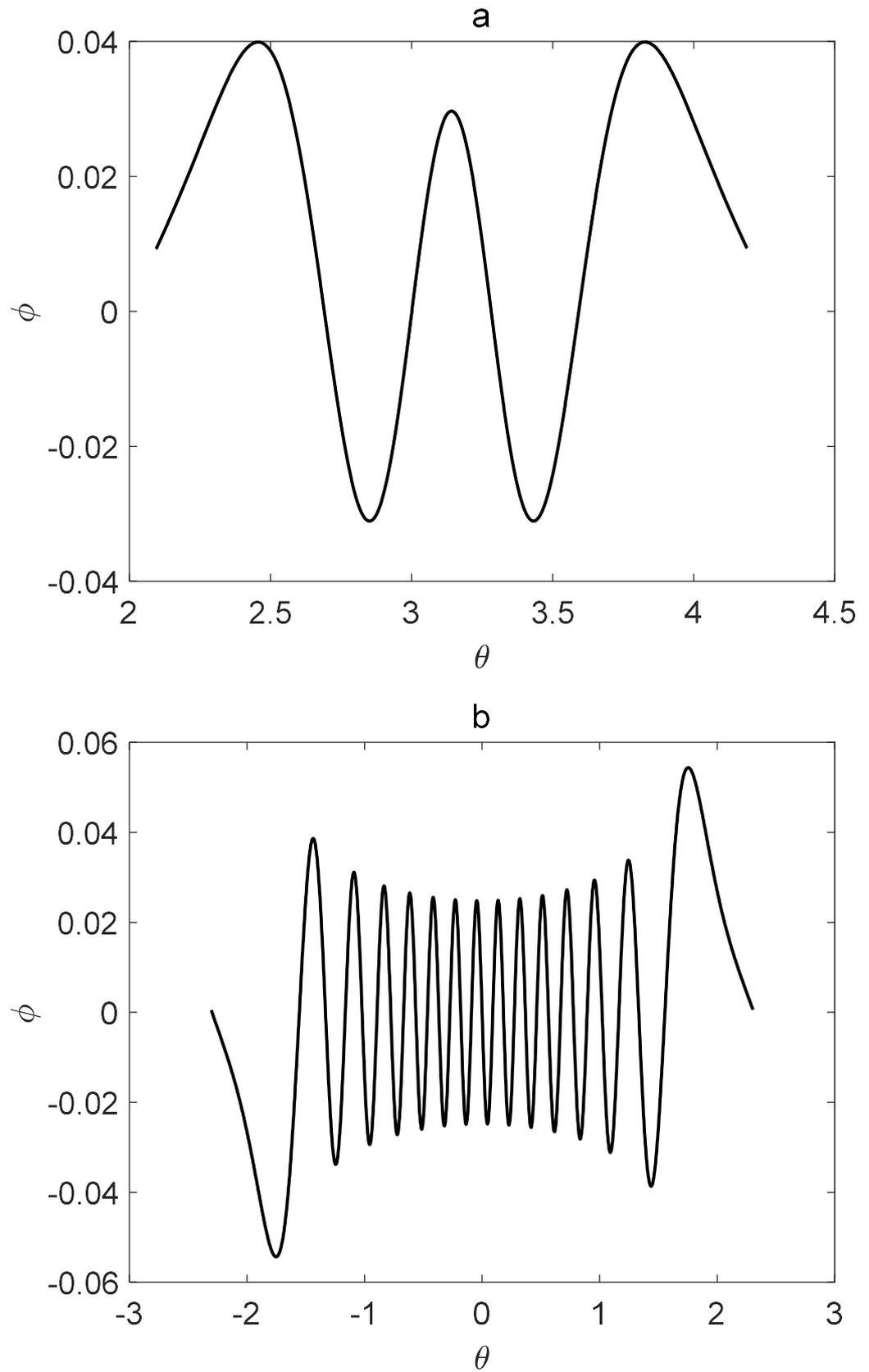

**Fig. 3.** Wave functions of the uppermost states in shallow well(a) and deep well (b).

Fig. 4 shows the result of a semiclassical calculation of the reduced phase $n' = S(2\pi)^{-1}$ as a function of energy, and its comparison with dependence (2.11) for free rotation. As the energy increases, an asymptotical convergence of the two curves occurs demonstrating an asymptotic behavior of the spectra.

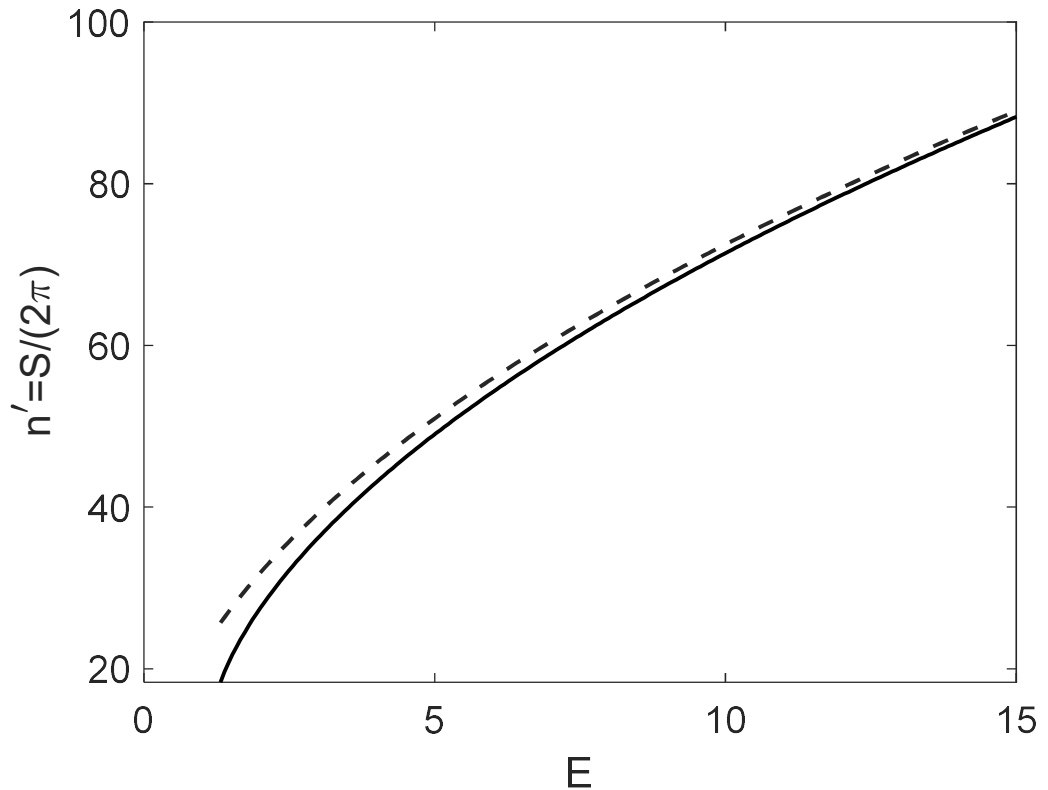

**Fig. 4.** Reduced phase as a function of energy for rotational motion of the Kapitza pendulum: solid line is the result of semiclassical approximation, dotted line is attributed to the free rotation.

## 6. Conclusions

The action of a high-frequency force on an inverted quantum linear oscillator leads to stabilization of the center of the wave packet, preserving its dispersion spreading [47]. In the Kapitsa quantum pendulum, the spreading of states is suppressed due to nonlinearity, and their destruction, without taking into account the stochastic thermal effect, can occur due to the resonant tunnel passage of the barrier separating the local minimum of the effective potential energy in the upper position of the pendulum from the global minimum in its lower position.

Comparison of the results obtained by different methods shows that the energy of the ground state, subject to stabilization, can be confidently calculated within the framework of the oscillator model with the first correction according to perturbation theory. For higher states, it is more adequate to use the semiclassical approximation or numerical calculations by the Numerov method. The semiclassical approximation reproduces the states not too close to the top of the potential, and its accuracy increases with the number of levels. A significant discrepancy between numerical calculations and simple analytical model estimates arises near the top of the potential barrier, where the standard WKB method is violated, and an additional level appears in comparison with the predictions of the semiclassical approximation. A detailed calculation of the states near the top of the potential barrier requires separate consideration and presupposes the knowledge of more accurate wave function in this energy range. This consideration can be done, for example, by using an inverted oscillator potential approximation near the top of the effective potential [43, 48]. Such an allowance is especially important for shallow well with one weakly bound state, when the usual semiclassical approximation is inapplicable.

In conclusion, we note that the considered problem of the motion of a system localized on a circle is close to the problem of the motion of a particle in a periodic waveguide. However, for the one-dimensional waveguide, the energy spectrum, which is found by the Floquet method, has, in the presence of resonance states, the structure of allowed energy bands, the centers of which in the range of vibration states coincide with the energies calculated for a circle motion. As to the quantum motion over the barrier, it will be characterized by a continuous energy spectrum.